\documentclass[12pt,epsfig]{article}
\usepackage{graphicx} 
\newcommand{\beq}{\begin{equation}}
\newcommand{\eeq}{\end{equation}}
\newcommand{\bea}{\begin{eqnarray}}
\newcommand{\eea}{\end{eqnarray}}
\newcommand{\be}{\beq}
\newcommand{\ee}{\eeq}
\begin{document}
\thispagestyle{empty}
\vspace*{-10mm}

\begin{flushright}
UND-HEP-04-BIG\hspace*{.08em}05\\
hep-ph/0411138 \\
\end{flushright}
\vspace*{4mm}

\centerline{\large\bf  `Per Aspera Ad Astra' \footnote{"Through the roughs to the stars"} -- }
\centerline{\large\bf A Short Essay on the Long Quest for CP Violation \footnote{To appear in the 
proceedings of `Time and Matter -- An International Colloquium on the Science of Time', Venice, Italy, August 11-17, 2002
}}
\vskip 0.3cm 
\centerline{I. I. Bigi$^a$} 

\vskip 0.3cm 
\centerline{$^a$Dept. of Physics, University of Notre Dame du Lac, Notre Dame, IN 46556, U.S.A.}
\centerline{e-mail: ibigi@nd.edu}
\vskip 1.0cm

\centerline{\bf Abstract}
After briefly explaining the special role played by violations of CP and T invariance and their connection with the baryon number of the Universe, I sketch the history of CP violation studies since 
its totally unexpected discovery in 1964. For about 30 years CP violation could be described by a single number; 
this has changed dramatically in  the years around the turn of the millenium: (i) The existence of 
{\em direct} CP violation was unequivocally established in the decays of long lived kaons. 
(ii)  For the first time CP violation was observed in a system other than that of neutral kaons, namely in 
$B\to \psi K_S$. The findings are in impressive agreement with the prediction of the CKM ansatz, which thus has 
been promoted to the status of a tested theory. These new insights were made possible by close feedback between 
theory and experiment as well as 
advances in detector design and a novel machine concept, namely that of an asymmetric collider. We also have 
direct {\em experimental} evidence that the observed CP violation in $K_L$ and $B$ decays is matched by 
a violation of microscopic time reversal violation, as required by CPT symmetry. 
More recently CP violation has been observed also in $B\to \pi^+\pi^-$ and $B\to K^{\mp}\pi^{\pm}$. A few comments are added on subtle aspects of direct CP violation. 
While we know that the CKM dynamics 
are irrelevant for generating the baryon number of the Universe -- i.e. hitherto unknown forces have to be driving it -- 
we have also learnt that such `New Physics' is likely to contain CP violation of sufficient strength.


\noindent {\bf Prologue}

\noindent The conference `Time and Matter' has as subtitle `An International Colloquium on the Science of Time'. 
The tale of the physicists' quest for CP violation fits naturally into this frame, since it presents us with several variations on the theme of {\em time}: 
the time it took to perform the experimental studies, the uneven rate of progress in our understanding, the time that had to be measured to reveal the sought-after CP asymmetry and the preference Nature shows on the microscopic level for the flow of time. It involves glorious applications of fundamental quantum mechanics, its superposition principle and of 
EPR correlations \cite{EPR} 
with their effects building up over macroscopic distances of centimeters, meters and even hundreds of meters. Finally it connects the "heavens" to the "earth" in that it provides us with a scenario where the seeds for the preponderance of matter over antimatter observed today can be generated {\em dynamically} in the very early Universe.

\section{On the Special Role of CP Violation}
\label{INTRO}

There are three {\em discrete} transformations of general interest, namely parity P, {\em microscopic} time reversal T (operationally amounting to reversal of motion $\vec p \to - \vec p$) and charge conjugation C, which replaces particles by their antiparticles. Originally it had been assumed without much reflection that all three represent symmetries of nature, 
since they were known to be conserved by the strong and electromagnetic forces. The first to fall from this pedestal were P and C. The 1957 discovery of P (and subsequently also of C) being violated by the weak forces did cause a paradigm shift. It was, however, realized that even {\em maximal} parity violation -- meaning there are left-, but {\em no} right-handed neutrinos -- does 
not necessarily imply  that nature exhibits a genuine preference for left over right. For  while the decay 
$\pi ^+ \to \mu^+ \nu_L$ produces only left-handed neutrinos, the antiparticle decay $\pi^- \to \mu^- \bar \nu_R$ yields 
right-handed neutrinos. They are referred to as antineutrinos, but at this point what is called particle and 
antiparticle is {\em pure convention}. For CP transformations relate the two processes; as long as CP invariance holds, 
they exhibit identical rates, and "left" and "right" is defined in terms of what one calls "positive" or "negative". This is  reminiscent of the definition "the thumb is left on your right hand" -- which is as correct as it is circular and thus useless. 

The observation of CP violation in 1964 by the Fitch-Cronin experiment \cite{FITCH} then came as another shock. 
\footnote{To my knowledge only Okun had stated explicitly {\em before} the Fitch-Cronin experiment 
that the question of CP 
invariance is one to be decided by experiment rather than theory.} For 
\be  
\frac{\Gamma (K_L \to \mu^+ \nu \pi^-)}{\Gamma (K_L \to \mu^- \bar \nu \pi^+)}\simeq 1.006 \neq 1 \; , 
\label{POSITIVE} 
\ee 
which is related to $K_L \to \pi \pi$ \cite{CPBOOK}, allows distinguishing a positive charge from a negative one through observation rather than convention. The discovery of CP violation thus changes our picture of Nature's structure even more profoundly than that of parity violation. At the same time it is quite `frustrating' that a CP invariant -- i.e. matter-antimatter symmetric -- 
world is such a `near-miss' with the difference on the $10^{-3}$ level, Eq.(\ref{POSITIVE}), in contrast to `maximal' 
parity violation: $\Gamma(\pi ^+ \to \mu^+ \nu_R)/\Gamma(\pi ^+ \to \mu^+ \nu_L)=0$. 

There are more features singling out CP violation as particularly special and more fundamental than parity violation: 

\noindent {\bf (i)} Almost any Lorentz invariant local quantum field theory has to possess CPT invariance. CP violation thus has to be matched by a commensurate T violation. I.e., nature distinguishing `left' and `right' implies her to do likewise between a `forward' and `backward' flow of time already on the {\em microscopic} level {\em beyond} the macroscopic statistical consideration expressed through thermodynamics' second law of entropy increase.  

\noindent {\bf (ii)} 
The leading contribution to the CP asymmetry in kaon decays involves $K^0 - \bar K^0$ oscillations, which represent quantum mechanical interference on a {\em macroscopic} scale. This allows the accurate measurement of truly tiny effects. The deviation from unity in Eq.(\ref{POSITIVE}) is driven by a tiny difference in the off-diagonal element of the 
$K^0 - \bar K^0$ mass difference: 
\be 
{\rm Im}M_{12} \simeq 1.1 \cdot 10^{-8} \; {\rm eV} \; \; \; \Leftrightarrow \; \; \;  
\frac{{\rm Im}M_{12}}{M_K} \simeq 2.2 \cdot 10^{-17} 
\ee
CP violation can thus be seen as the smallest observed (rather than hypothesized) violation of a symmetry. In turn this means that searches for CP violation in different systems serve as very high sensitivity probes for hitherto unknown physics. 

\noindent {\bf (iii)} 
In a 1967 paper \cite{SAKH} the famous Russian physicist (and dissident) Andrei Sakharov noted   that the existence of CP violation opened the path towards a new paradigm with a cosmic connection, namely to  understand the baryon number of the Universe, for which data yield $\sim 10^{-9}$, as a {\em dynamically generated} number rather than an arbitrary initial value \footnote{By baryon number of the Universe we mean effectively the ratio of the number of baryons in the Universe relative to that of photons in the microwave background radiation; we know of no significant source of {\em primary}  antibaryons in the Universe.}; this number can be qualitatively characterized through the following dual phrase: while the 
Universe is not empty, it is almost empty. 

To achieve this goal, three requirements have to be met: 
\begin{itemize}
\item 
There are forces changing the baryon number; 
\item 
CP invariance is broken; 
\item 
the Universe is out of thermal equilibrium.  
\end{itemize}
 The probably most ambitious goal in CP studies is to identify this cosmic connection, i.e. to find other manifestations of the CP violating forces that drive the baryon number of the Universe, which could be probed in reproducible laboratory 
experiments. 

\noindent {\bf (iv)} A particularly subtle feature is the following \cite{CPBOOK}. Since the 
time reversal operator T is {\em anti}unitary, we have $T^2 = \pm 1$. Thus the Hilbert space for systems invariant under 
time reversal consists of two disjoint sectors, one with $T^2 = +1$ and the other with $T^2 = -1$. Consider an energy eigenstate $|E\rangle$ belonging to the latter; it can be shown that the degenerate state $T|E\rangle$ is {\em orthogonal} to $|E\rangle$, 
which is referred to as `Kramers' degeneracy' \cite{KRAMERS}. This implies that $|E\rangle$ carries an internal degree of freedom that is changed by T, {\em if} $T^2=-1$. One should note that this property was derived 
{\em without} any reference to spin, half-integer or otherwise! It is of course completely consistent with the transformation 
behaviour of spin degrees of freedom. This conceptual consequence of Kramers' degeneracy can be rephrased as follows: a world where all systems have $T^2 = +1$ would be conceivable -- as is one with only integer-spin or only 
with states obeying Bose-Einstein statistics; yet once again nature reveals its clear tendency to realize dynamical structures that are mathematically admissible -- and to do it in a very efficient way: odd-integer states double as fermions and as systems  with $T^2 = -1$. There is a more practical consequence of Kramers' degeneracy as well: an odd-number electron system placed inside an external electrostatic field will always exhibit (at least) a {\em two-}fold degeneracy, no matter how complicated that field is; this property does {\em not} hold for an even-number electron system. 

While the Fitch-Cronin result was initially greeted with surprise, dismay and even shock, attempts at `denial' 
\footnote{At first there were suggestions to make the observation of $K_L \to \pi\pi$ compatible with CP symmetry by either introducing {\em non}linear terms into the Schr\" odinger equation thus abandoning the superposition principle of 
quantum mechanics \cite{ROOS} or by postulating that actually $K_L \to \pi \pi U$ had occurred with a hypothetical neutral pseudoscalar particle U having escaped detection. Additional data sealed off these escape routes.} 
were soon abandoned. It was realized that novel insights and perspectives onto Nature's Grand Design could be gained by continuing a dedicated and comprehensive study of CP violation in a wide array of different reactions. The ensuing story is a fascinating one. It can be seen as describing the high energy physics paradigm in a nutshell: a fundamental question is at stake; long periods of {\em seeming} stagnation are often followed by intervals of unexpected twists and turns, even breakthroughs; the conclusion of one chapter often comes with the first message from the next chapter; progress is achieved through the interplay and constructive interference of theory, experiment and new technologies in turn taking the lead. The outcome is one where participants can take a great deal of pride and others feel a great deal of envy. 

\section{1964 - 1998: The Long Wait} 
\label{1998} 

On the surface the long period between just after the discovery of CP violation in 1964 and 1998 appeared one of stagnation; in retrospect it can be seen rather as one of fermentation, with many things of future importance developing 
just below the surface.  

\subsection{Experimental Searches}
\label{EXPSEARCH}

After a long period of dedicated and ingenious experimentation described in more detail in the talks by Blucher 
\cite{BLUCHER} and Zavrtanik \cite{ZVARTANIK} CP violation had been found in three classes of processes all starting from neutral kaons: (i) $\Gamma (K_L \to \pi \pi) \neq 0$; 
(ii) $\Gamma (K_L \to l^+\nu \pi^-)\neq  \Gamma (K_L \to l^-\nu \pi^+)$; 
(iii) rate$(K^0(t) \to \pi \pi)$ $\neq$ rate$(\bar K^0(t) \to \pi \pi)$. Assuming CPT invariance one finds that these transitions 
can be described by two ratios of transition amplitudes: 
\be 
\eta _{+-} \equiv \frac{T(K_L\to \pi^+\pi^-)}{T(K_S\to \pi^+\pi^-)} \; , \; 
\eta _{00} \equiv \frac{T(K_L\to \pi^0\pi^0)}{T(K_S\to \pi^0\pi^0)}
\ee
Using the notation 
\be 
\eta_{+-} = \epsilon + \epsilon^{\prime}\; , \; \eta_{00} = \epsilon - 2 \epsilon^{\prime}
\ee
one finds that $\epsilon \neq 0$ describes `indirect' CP violation, i.e. CP violation in the 
$\Delta S=2$ dynamics driving $K^0 - \bar K^0$ mixing, which prepares the initial $K_L$ state and 
$\epsilon^{\prime} \neq 0$ `direct' CP violation in the $\Delta S=1$ decay sector. Up to 1998 all data could 
be described by a single non-vanishing real number, namely $|\epsilon|$ 
(or arg$(M_{12}/\Gamma_{12})$). 
The situation was quite unsettled concerning {\em direct} CP violation: 
\be 
{\rm Re} \frac{\epsilon ^{\prime}}{\epsilon _K} = 
\left\{   
\begin{array}{ll}
(2.3 \pm 0.65) \cdot 10^{-3} & NA\; 31 \\
(1.5 \pm 0.8) \cdot 10^{-3} & PDG \; '96\;  {\rm average} \\
(0.74 \pm 0.52 \pm 0.29) \cdot 10^{-3} & E\; 731 \\
\end{array} 
\right. 
\label{DIRECTCPDATA} 
\ee 
The CPLEAR experiment had performed the socalled Kabir Test of T invariance and under very general assumptions found an asymmetry commensurate 
with the observed CP violation as predicted by CPT invariance \cite{ZVARTANIK}: 
\be
\frac{\Gamma (K^0 \Rightarrow \bar K^0) - 
\Gamma (\bar K^0 \Rightarrow K^0)} 
{\Gamma (K^0 \Rightarrow \bar K^0) + 
\Gamma (\bar K^0 \Rightarrow K^0)} = 
(6.3 \pm 2.1 \pm 1.8) 
\cdot 10^{-3} \; \; \; 
CPLEAR 
\label{CPLEAR}
\ee 
 
Truly impressive sensitivities had been achieved in searches for T violation through electric dipole moments for neutrons 
and electrons, respectively \cite{CPBOOK}: 
\bea
d_N &<& 6.3 \cdot 10^{-26} \; ecm \; \; \; {\rm from \; ultracold \; neutrons} \\
d_e &=& (-0.3 \pm 0.8) \cdot 10^{-26} \; ecm \; \; \; {\rm from \; atomic \; EDM} 
\eea
I.e., the upper bound on $d_N$ amounts to a shift 12 orders of magnitude smaller than the radius of the neutron. 
This corresponds to searching for a shift of less than the width of human hair in an object the size of the earth !

\subsection{Theoretical Models}
\label{THMODEL}

The `superweak' ansatz positing that CP violation resides only in $\Delta S=2$ transitions -- and thus 
$\epsilon /\epsilon ^{\prime} = 0$ -- was put forward by Wolfenstein already in 1964 
\cite{SUPERWEAK}. Yet it has to be kept in 
mind that it constitutes a {\em classification} scheme rather than a {\em dynamical} model, let alone a theory. The 
community might be forgiven for not worrying unduly over a tiny effect -- characterized by 
BR$(K_L \to \pi^+\pi^- )\simeq 2.3 \cdot 10^{-3}$ -- at a time where there was no renormalizable theory for the 
weak forces, and one had to deal with {\em infinities} in decay widths. However it is quite remarkable that after the 
emergence of a renormalizable theory in the late 1960's -- namely the Standard Model (SM) based on 
$SU(2)_L \times U(1)$ gauge interactions -- it was not realized for several years that New 
Physics, i.e. physics beyond the standard model {\em of that time} had to exist. 
This lack of a theory was stated unequivocally in the 1973 paper by 
Kobayashi and Maskawa \cite{KM}. They also listed the various scenarios that can support CP breaking (while maintaining 
CPT invariance): right-handed charged currents \footnote{Mohapatra had suggested this option already in 1972 
\cite{MOHA}.}, extra Higgs fields or -- the existence of (at least) a third quark family.  This last of their options is now referred to as the KM description\footnote{The only option they missed is the possibility that a nonabelian gauge theory like QCD can break CP 
invariance through topological effects in its ground state \cite{CPBOOK}.}.  

They made their suggestions when neither right-handed currents nor any Higgs states had been found -- a fact that 
still holds today -- and when only three quark `flavours' were known, namely $u$, $d$ and $s$; i.e. even the second family was not complete\footnote{Kobayashi and Maskawa had benefitted in two ways from the `genius loci' of Nagoya University where they both worked at that time: (i) While the notion of quarks as truly dynamical objects rather than convenient mathematical entities had not been universally accepted, this was not doubted at Nagoya, which was the birth place of the Sakata model. 
(ii) Niu, a prominent professor at Nagoya, had found evidence for a charm hadron in his cosmic ray data in 1971 
\cite{NIU}. At Nagoya it was thus `known' that at least two full families of quarks existed in nature.}! 

Kobayashi and Maskawa extended an earlier observation by Cabibbo that {\em mass} eigenstates are not necessarily 
{\em interaction} or {\em `flavour'} eigenstates as well. Then there are nontrivial transformations 
$T_L^U$ and $T_L^D$ relating the left-handed flavour 
eigenstates of $U=u,c,...$ and $D=d,s,...$ quarks, respectively, to mass eigenstates; the charged current couplings of pairs of 
$U$ and $D$ quarks are described by a matrix, the CKM matrix 
\be
V^{CKM} = T_L^U(T_L^D)^{\dagger}   
\ee
This matrix has to be unitary as long as the weak interactions are described by a {\em single} $SU(2)$ gauge theory. 
Kobayashi and Maskawa pointed out that with two families only -- $(u,d)$ and $(c,s)$ -- $V_{CKM}$ cannot contain a 
{\em physical} complex phase, which is required to implement CP violation. Yet for three families, where one has a 
$3\times 3$ matrix, it contains three mixing angles -- in analogy to the Euler angles of rotation matrices -- 
plus one complex phase, the KM phase $\phi  _{KM}$. 

This observation has lead to enhanced visibility of basic trigonometry in fundamental physics. The unitarity of the $3\times 3$ CKM matrix implies two classes of constraints: 
\be 
\sum_{l=1}^3
(V_{il}^{CKM})^*V_{lj}^{CKM} = \delta_{ij}  
\label{UNITARITY}
\ee
For $i=j$ this represents `weak universality' (for example: 
$|V_{ud}|^2 + |V_{us}|^2 + |V_{ub}|^2 =1$) , for $i\neq j$ a {\em triangle} relation in the complex plane. There are actually 
six triangles: they are of very different shapes, yet share one important feature: they all have the same area, which can 
be expressed through the so-called Jarlskog variable \cite{JARL}: 
\be 
{\rm area} = \frac{1}{2} J \; , \; \; J = |{\rm Im}V^*_{km}V_{lm}V_{kn}V^*_{ln}| 
\ee 
irrespective of the indices $k$, $l$, $m$, $n$. This feature of equal areas reflects the fact that there is a single 
irreducible complex phase for three families. The angles in these 
triangles control CP asymmetries in the different charged current transitions like the decays of kaons and $B$ mesons, and they are all driven by the KM phase $\phi _{KM}$.  
The main point here is that the CKM description involves a {\em high degree of overconstraints}: 
\begin{itemize}
\item 
The angles which control the CP asymmetries can be determined by the sides of the triangles, which in turn can be inferred from CP {\em in}sensitive rate measurements. 
\item 
The different angles are related to each other, and all depend on $\phi_{KM}$. 

\end{itemize}

The KM ansatz naturally predicts direct CP violation, i.e. it does {\em not} represent a superweak description.

\subsection{The Distinct Attraction of Beauty}
\label{ATTRACTION}

According to the KM ansatz CP violation is due to the interplay of (at least) three quark families. Kaons made up from 
quarks of the first two families ($d,u$ and $s$) are sensitive to the third family due to quantum corrections, yet those are suppressed, 
since the quarks of the third family are so much heavier; this provides a natural qualitative explanation, why CP violation 
is so feeble in kaons, i.e. why CP invariance is such a `near miss' in strange decays. 

The situation changes very significantly for the decay of $B$ mesons, since the $b$ quark already belongs to the third family. Right after the first evidence for $b$ quarks was found, it was recognized that $B$ decays had the potential to exhibit sizable CP asymmetries. In 1980 it was realized how this potential could be tapped \cite{BS1}. 
More specifically it was pointed out that some CP asymmetries in $B$ decays could be two orders of magnitude larger than what had been found in $K_L \to \pi \pi$ -- even approaching 100 \% -- and that they could be related reliably to basic parameters of the CKM description.  This applies in particular to the mode $B_d \to \psi K_S$. The CP asymmetry was predicted to exhibit another striking signature beyond its size, namely a peculiar dependence on the time of decay 
$t$ 
\be 
{\rm rate} (B_d(t)[\bar B_d(t)] \to \psi K_S) \propto e^{-t/\tau _B} (1- [+] A {\rm sin}\Delta m_B t) \; , 
\label{ASYM}
\ee
since it involves $B_d - \bar B_d$ oscillations in an essential way. The asymmetry parameter $A$ for this transition 
can reliably be expressed through one angle of the unitarity triangle, see Fig.\ref{TRIANGLE}:
\be 
A = {\rm sin}2\phi_1
\label{PHI1}
\ee

The final state $\psi K_S$ does not 
reveal whether it came from a $B_d$ or $\bar B_d$ decay; that principal ambiguity is actually essential for the 
asymmetry in Eq.(\ref{ASYM}) to arise, since the asymmetry is due to the interference of two {\em coherent} 
amplitudes. Thus one needs {\em independent} information on the `flavour' identity 
of the decaying $B$ meson; this can be achieved by `associated' production, as sketched later. 

At first not much attention was paid to these suggestions, partly because the lifetime $\tau _B$ of $B$ mesons 
was not known, let alone $\Delta m_B$, the rate of $B^0 - \bar B^0$ oscillations. A highly significant change in perception occurred when first the $B$ lifetime was measured and found to be surprisingly "long" at ${\cal O}(psec)$ followed by the 
observation of $B^0 - \bar B^0$ oscillations with $\Delta m_B \sim 0.7/ \tau_B$. These observations established a new paradigm: 
\begin{itemize}
\item 
It revealed a peculiar hierarchical pattern in the CKM matrix most transparently expressed through the Wolfenstein 
representation:
$$
V^{CKM} \equiv  
\left( 
\begin{array}{ccc} 
V_{ud} & V_{us} & 
V_{ub} \\
V_{cd} & V_{cs} & 
V_{cb}\\ 
V_{td} 
& V_{ts}& V_{tb}
\end{array}
\right)
\nonumber 
$$
\be 
\simeq 
\left( 
\begin{array}{ccc} 
1 - \frac{1}{2} \lambda ^2 & \lambda & 
A \lambda ^3 (\rho - i \eta + \frac{i}{2} \eta \lambda ^2) \\
- \lambda & 1 - \frac{1}{2} \lambda ^2 - i \eta A^2 \lambda ^4 & 
A\lambda ^2 (1 + i\eta \lambda ^2 ) \\ 
A \lambda ^3 (1 - \rho - i \eta ) 
& - A\lambda ^2 & 1 
\end{array}
\right) 
\label{WOLFKM} 
\ee
\item 
This pattern 
lead to the realization there is a unitarity triangle where all three sides are of order $\lambda^3$ in length; accordingly all three angles are naturally large. This special triangle, shown in Fig. \ref{TRIANGLE}, describes transitions of $B$ 
mesons. Thus one predicts CP asymmetries of several$\times$ 10\% \cite{BS2}. 

\begin{figure}[h]
\centerline{
\includegraphics[height=5cm]{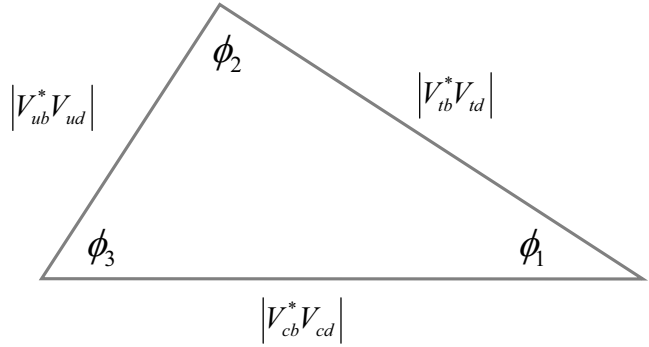}}
\caption{The CKM Unitarity Triangle for $B$ decays
\label{ckm}}
\end{figure}



\item 
It just happened that the detector technology for tracking decay times $\sim$ 1 psec was available 
`off the shelf', since it had been 
developed for prior studies of charm hadrons. 
\item 
Events with times of decay much shorter than 1 psec, which largely escape tracking, contribute little to the asymmetry in Eq.(\ref{ASYM}). 
\end{itemize} 

There are various experimental setups for undertaking such measurements that have been and will be pursued 
in the future \cite{OPAL,CDF}, in particular at hadronic colliders. Yet the cleanest experimental stage is provided by colliding electron and 
positron beams: 
\be 
e^+e^- \to \Upsilon (4S) \to B_d \bar B_d 
\label{UPS}
\ee
For the $\Upsilon (4S)$ resonance yields an enhanced production rate for $B$ mesons,  
and all final state particles are decay products of one or the other $B$ meson. 

There is another equally important point, which at first represents a serious challenge. To {\em experimentally define} a CP asymmetry, one can {\em flavour tag} an event by observing a flavour 
specific transition like $B_d \to l^+\nu X$ or $\bar B_d \to l^- \nu X$ 
{\em in conjunction} with the {\em non-}specific mode 
$B_d/\bar B_d \to \psi K_S$ to compare 
\be 
B_d \bar B_d \to (l^+X)_B(\psi K_S)_{B/\bar B} \; vs. \; B_d \bar B_d \to (l^-X)_{\bar B}(\psi K_S)_{B/\bar B} 
\ee
Denoting the times of the two decays by $t_1$ and $t_2$, respectively, one finds 
\be 
{\rm rate}(B_d(t_1) \to l^+\nu X),\bar B_d(t_2)\to \psi K_S) \propto 
e^{-(t_1 + t_2)/\tau_B}[1- A{\rm sin}\Delta m_B(t_1 - t_2)] \; ; 
\label{COMBCP} 
\ee
likewise for ${\rm rate}(\bar B_d(t_1) \to l^-\nu X), B_d(t_2)\to \psi K_S)$. 
The relative minus sign between $t_1$ and $t_2$ is due to the fact that the $B\bar B$ pair in 
$\Upsilon (4S) \to B_d \bar B_d$ forms a C {\em odd}  configuration. 
There was a significant fly-in-the-ointment, though: One cannot measure the time of decay {\em directly}.  
Silicon microvertex detectors instead allow to identify the {\em location} of the {\em decay} vertex. 
Knowing the {\em production} vertex as well 
one can infer the distance the $B$ meson traveled; from it and the $B$ momentum one obtains the lifetime. Space 
distances are thus translated into time intervals. However the $\Upsilon (4S)$ is barely above the 
$B_d \bar B_d$ production threshold; therefore the $B$ mesons move slowly in the 
$\Upsilon (4S)$ rest frame. The distance they cover before decaying is too short to be resolved by existing detectors. 
Yet integrating Eq.(\ref{COMBCP}) over all times $t_{1,2}$ removes the asymmetry term. I.e., a search for such a CP asymmetry would amount merely to a test for . 

An intriguing answer to this challenge would utilize quantum mechanical coherence as follows: 
in the reaction 
\be
e^+e^- \to B_d^* \bar B_d + h.c. \to B_d \bar B_d \gamma \; . 
\ee 
The $B\bar B$ pair now forms a C {\em even} state; accordingly the time difference $t_1 - t_2$ in the 
sin$\Delta m_B(t_1 - t_2)$ term of Eq.(\ref{COMBCP}) is replaced by the sum $t_1 + t_2$; 
integrating over all time $t_1$, $t_2$ 
then yields a non-vanishing contribution. 
I.e., one then could search for a CP asymmetry in $B_d \to \psi K_S$ {\em without} resolving decay vertices. 
Yet such an undertaking suffers from a lack of the necessary  statistics. 

Oddone \cite{ODDONE} suggested to cut the Gordian knot for $e^+e^- \to \Upsilon (4S) \to B_d \bar B_d$ by using 
{\em asymmetric} colliding 
beams, which makes the $\Upsilon (4S)$ move in the lab frame. With a sufficiently large beam energy asymmetry the $B$ mesons receive a boost in the lab frame making them travel a distance long enough to be resolved. 

This was quite an innovative and daring proposal, since there was no experience with asymmetric colliders; furthermore 
integrating two separate beam lines of greatly different energies -- and high intensity on top of that -- into a detector 
posed novel problems. Yet it was a challenge that was thought, at least by some adventurous souls, worth the effort. 

This is best demonstrated by the fact that a race ensued to build an asymmetric high intensity $e^+e^- \to \Upsilon (4S)$ collider together with a detector capable of operating in a high radiation environment. It was joined by two teams in the late 1980's: the BABAR collaboration  based at SLAC, the Stanford Linear Accelerator Center, in the USA, and the BELLE collaboration based at KEK, the Japanese High Energy Accelerator Research Organization. While the designs of the two teams for the detectors and accelerators differed in many details -- like energy asymmetries, injection systems for the beams, luminosities, particle id for the detectors etc.-- both were truly ambitious: the SLAC project aimed 
for a luminosity of 
$3\cdot 10^{33} \; cm^{-2}s^{-1}$ and KEK even for $10^{34} \; cm^{-2}s^{-1}$, when the $10^{33}$ level had never been achieved before even with conventional colliders! These design intensities amount to the production of 
about three and ten $B\bar B$ pairs per second, respectively.

\subsection{On the Eve of a `Phase Transition'}
\label{EVE}

In 1998 the  theoretical status of CP violation could be summarized as follows:: 
\begin{itemize}
\item 
The strength of indirect CP violation observed in $K_L$ decays could be reproduced within the CKM description without forcing any parameter outside the range inferred for it from other measurements. 
\item 
There appeared to be more consensus about the strength of {\em direct} CP violation among the theory predictions than between the two sets of data from NA31 and E731, Eq.(\ref{DIRECTCPDATA}): most authors predicted 
Re($\epsilon^{\prime}/\epsilon$) not to exceed $10^{-3}$, while some heretics --early ones \cite{EARLY} 
and `just in time' ones predicted larger values: 
$(1.7 ^{+1.4}_{-1.0} )\cdot 10^{-3}$ 
\cite{TRIESTE}. 
\item 
The KM ansatz predicted unobservably small values for electric dipole moments of neutrons, electrons and atoms, 
i.e. below $10^{-30}\; ecm$. 
\item 
There is one caveat: it had been realized that QCD, the nonabelian gauge theory of the strong interactions, does 
{\em not} conserve CP invariance {\em naturally}. The nontrivial topological structure of its ground state induces a P and T 
{\em odd} term in QCD's effective Lagrangian with an a priori unknown coefficient $\theta _{QCD}$, for which the 
`natural' expectation is $\theta _{QCD} \sim {\cal O}(1)$. This term would generate an EDM for the neutron; the 
{\em non}observation of the later imposes $\theta_{QCD} \leq {\cal O}(10^{-9})$ -- i.e. smaller by many orders of magnitude than the `natural' expectation! This means that either the neutron's EDM could be `just around the corner' -- i.e. any improvement 
in experimental sensitivity might reveal an effect -- or again  unobservably small, since `natural' explanations of 
$\theta_{QCD} \leq {\cal O}(10^{-9})$ based on a Peccei-Quinn symmetry drive $\theta_{QCD}$ down by several additional orders of magnitude \cite{CPBOOK}.  
\item 
The CKM description actually {\em post}dicted $\epsilon$ (as it did for the $K_L$-$K_S$ mass difference $\Delta m_K$), 
and its predictions for other observables like $\Delta m_B$ suffer from considerable uncertainties. Nevertheless it was argued \cite{CPBOOK} that the ability of the CKM scheme to accommodate a body of observables spanning six or seven orders of magnitude in energy has to be seen as highly nontrivial, in particular since it was achieved with fundamental quantities 
like quark masses and the CKM parameters shown in Eq.(\ref{WOLFKM}) that would have seen frivolous -- if not forced upon us by the data. Thus some of us had considerable confidence that the CKM prediction for a large CP asymmetry 
in $B_d (t) \to \psi K_S$ would be confirmed. 

This confidence was expressed by saying that there was no `plausible deniability' for the KM ansatz, if no large CP asymmetries were found in $B$ decays; in 1991/92, i.e. before the top quark was discovered and its mass measured directly, 
the expectation \cite{BEFORETOP} 
\be 
{\rm sin}2\phi_1 \sim 0.6 \div 0.7 
\label{PHI1PRED1}
\ee
was formulated, and in 1998 an even more specific prediction was given \cite{PRECISE}: 
\be 
{\rm sin}2\phi_1 \sim 0.72 \pm 0.07 
\label{PHI1PRED2}
\ee
The point of listing this last number is to emphasize that it {\em was made} as a true 
{\em pre}diction rather than to endorse the error quoted there. It should be added that us Bavarians always 
admire courage, in particular of the somewhat reckless kind. 
\item 
The first suggestion of sizable CP asymmetries in $B$ decays like $B\to K \pi$ was actually made by the authors of Ref.\cite{SONI}. A re-analysis by the authors of Ref.\cite{JARL2} lead to 
\be 
A_{CP}^{K\pi} \equiv \frac{\Gamma (\bar B^0 \to K^- \pi^+)- \Gamma (B^0 \to K^+ \pi^-)} 
 {\Gamma (\bar B^0 \to K^- \pi^+)+ \Gamma (B^0 \to K^+ \pi^-)} = - 0.10
 \label{BKpi}
\ee
as a reasonable, though not firm prediction.

\end{itemize}

 \section{The `Phase Transition' at the Turn of the Millenium}
 \label{PHASE}
 
 During a relatively short time interval around the year 2000 data provided us with several seminal insights 
 answering old questions -- and raising new ones. 
 
 \subsection{The Conclusion of an Epoch}
 \label{ENDEPOCH}
 
 In 1999, after more than thirty years of dedicated and ingenious experimentation \cite{BLUCHER}, 
 KTEV \cite{KTEV} and NA48 \cite{NA48} {\em conclusively} confirmed earlier evidence from NA31 that indeed there is {\em direct} CP violation 
 in $K_L$ decays.  The 2003 world average reads 
 \be
 {\rm Re} \frac{\epsilon ^{\prime}}{\epsilon} = (1.66 \pm 0.16) \cdot 10^{-3} \; \; 
 \hat = \; \; 
 \frac{\Gamma (K^0 \to \pi ^+\pi^-) - \Gamma (\bar K^0 \to \pi ^+\pi^-)}
 {\Gamma (K^0 \to \pi ^+\pi^-) + \Gamma (\bar K^0 \to \pi ^+\pi^-)} = (5.5 \pm 0.6) \cdot 10^{-6} 
 \label{WA03}
 \ee
 The second number even more than the first one indicates what kind of achievement lies behind these data. 
 The physicists involved in these experiments have earned our respect, and they certainly have my admiration. 
 Establishing direct CP violation for the first time is a discovery of the first rank irrespective of what theory does or does 
 not say. At the present status of our knowledge (or the lack thereof) it is not inconsistent with the CKM description. 
 Nature has exhibited its slightly more malicious side here, since $\epsilon^{\prime}$ receives several contributions with the two largest ones coming in with the opposite sign. Thus we cannot count on theory yielding a 
 {\em definitive} answer soon. Yet again, I find it highly nontrivial that theory yields the correct number to within a factor of two or so.

 \subsection{The Beginning of a New Era}
 \label{NEWERA}
 
 As already mentioned, the two $B$ factories had very ambitious goals concerning their luminosities and reliabilities. They have actually met them -- and surpassed them. BELLE and BABAR with design luminosities of 
 $1\cdot 10^{34}$ and $3\cdot 10^{33} cm^{-2}s^{-1}$, respectively, in 2004 have achieved running at 
 $1.2\cdot 10^{34} cm^{-2}s^{-1}$ and $8\cdot 10^{33} cm^{-2}s^{-1}$, respectively. They presented their first still inconclusive data on $B_d \to \psi K_S$ in 2000:  
\bea 
 {\rm sin}2\phi_1 &=& 0.45 \pm 0.44 \pm 0.09 \; \; \; {\rm BELLE \; '00} \\ 
 {\rm sin}2\phi_1 &=& 0.12 \pm 0.37 \pm 0.09 \; \; \; {\rm BABAR \; '00} 
 \label{PHI100}
\eea
In the summer of 2001 they established the existence of an 
asymmetry, the first one outside the decays of neutral kaons and a truly large one 
\footnote{To obtain these small errors various other flavour tagging modes beyond semileptonic decays had to be 
used and other final states like in particular $B_d/\bar B_d (t) \to \psi K_L$; for the latter one predicts an asymmetry of equal size, yet opposite sign to that in $B_d/\bar B_d(t) \to \psi K_S$.}: 
\bea 
 {\rm sin}2\phi_1 &=& 0.99 \pm 0.14 \pm 0.06 \; \; \; {\rm BELLE \; '01} \\ 
 {\rm sin}2\phi_1 &=& 0.59 \pm 0.14 \pm 0.05 \; \; \; {\rm BABAR \; '01} 
 \label{PHI101}
\eea
 Two years later the data had converged to an amazing degree: 
 \be 
  {\rm sin}2\phi_1 = \left\{ 
\begin{array}{lll} 
0.733 \pm 0.057  \pm 0.028 & \; \; \; {\rm BELLE \; '03} & \; \;   {\rm with} \sim 1.2 \cdot 10^8 \; B\bar B \\ 
0.741 \pm 0.067  \pm 0.030 & \; \; \; {\rm BABAR \; '03} & \; \;  {\rm with}  \sim 0.8 \cdot 10^8 \; B\bar B\\
0.736 \pm 0.049   & \; \; \; \;  {\rm world \;  average \; '03} & \\
0.726 \pm 0.037   & \; \; \; \;  {\rm world \;  average \; '04} & 
 \end{array}  
 \right.  
 \label{PHI103}
 \ee
 leading to the following general statements: the CP asymmetry in $B_d \to \psi K_S$ is there, 
 and it is huge, fully as expected, see Eq.(\ref{PHI1PRED2})
\footnote{The procession of these numbers reflects a better understanding of the detectors in addition 
to increasing statistics. It should also remind theorists to consider experimental uncertainties when interpreting data.}
!
 
 Hence I conclude: 
\begin{itemize}
\item 
The CKM paradigm has been promoted from an {\em ansatz} to a {\em tested theory}. 
\item  
CP violation has actually been `demystified': {\em if} the dynamics are sufficiently multilayered such that they can 
support CP violation (like the existence of at least three quark families), the latter can be truly large; i.e. there is no intrinsic reason why the complex phases should be small. 
\end{itemize}

\subsection{The Unsung Hero}
\label{HERO}
 
Hadronization -- the formation of hadrons out of quarks -- is usually listed as an unwelcome 
complication greatly impeding our description of CP violation, since the strong forces have not been brought under full theoretical 
control. The latter is certainly true -- yet so is the fact that hadronization greatly enhances the features of CP breaking and thus facilitates its observability through three effects: 
\begin{itemize}
\item 
The existence of pions and kaons with the latter only moderately above the three pion threshold reduces the rate for 
the CP conserving $K_L \to 3 \pi$ process relative to the CP violating 
$K_L \to 2\pi$ one by a factor close to $\Gamma (K_S)/\Gamma (K_L) \sim 500$. 
\item 
It awards `patience'; i.e., an initial beam of $K^0$ and $\bar K^0$ turns into a pure $K_L$ beam, since the $K_S$ component decays away quickly. 
\item 
CP violation can be established through the {\em existence} of a transition -- here $K_L \to \pi \pi$ -- rather than an 
asymmetry between two allowed processes. 

\end{itemize} 
Hadronization should thus be recognized as the hero of the tale of CP violation rather than the villain it is usually depicted.

 \subsection{EPR Correlations -- a Precision Tool Rather than a Paradox}
 \label{EPR}
 
 The BABAR and BELLE analyses are based on a glorious application of quantum mechanics and in 
 particular EPR correlations\cite{EPR}.  
 At first it would seem that an asymmetry of the form given in Eq.(\ref{ASYM}) could not 
 be measured for practical reasons. For in the reaction
 \be 
 e^+e^- \to \Upsilon (4S) \to B_d \bar B_d
 \label{UPS4S}
 \ee
 the point where 
 the $B$ meson pair is produced is ill determined due to the finite size of the electron and positron beam spots: the 
 latter amounts to about 1 mm in the longitudinal direction, while a $B$ meson typically travels only about a quarter 
 of that distance before it decays. 
 It would then seem that the length of the flight path of the $B$ mesons is poorly known and that 
 averaging over this ignorance would greatly dilute or even eliminate the signal. 
 
 It is here where the existence of a EPR correlation comes to the rescue. While the two $B$ mesons in the reaction 
 of Eq.(\ref{UPS4S}) oscillate back and forth between a $B_d$ and $\bar B_d$, they change their flavour identity in a 
 {\em completely correlated} way.  For the $B \bar B$ pair forms a C odd state; Bose statistics then tells us that there cannot be two identical flavour hadrons in the final state: 
 \be 
 e^+e^- \to \Upsilon (4S) \to B_d \bar B_d \not \to B_d B_d, \; \bar B_d \bar B_d
 \label{NOTID}
 \ee
 Once one of the $B$ mesons decays through a flavour specific mode, say $B_d \to l^+\nu X$ 
 [$\bar B_d \to l^- \bar \nu X$], then we know unequivocally that the other $B$ meson was a 
 $\bar B_d$ [$B_d$] at {\em that} time. The time evolution of $\bar B_d(t) [B_d(t)] \to \psi K_S$ as described by 
 Eq.(\ref{ASYM}) starts at {\em that} time as well; i.e., the relevant time parameter is the {\em interval between} 
 the two times of decay, not those times themselves. That time interval is related to -- and thus can be inferred from -- 
 the distance between the two decay vertices, which is well defined and can be measured.


\begin{figure}[h]
\centerline{
\includegraphics[height=11cm]{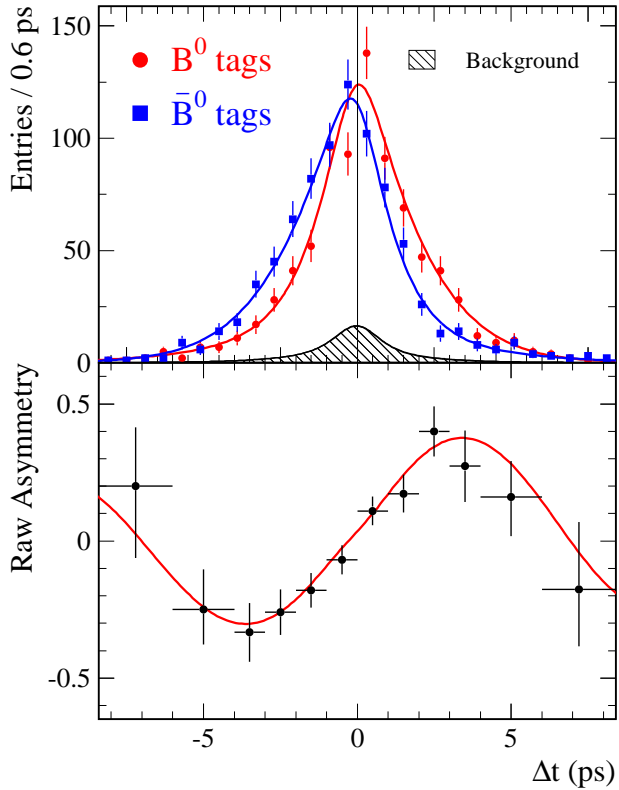}}
\caption{The observed decay time distributions for $B^0$ (red) and $\bar B^0$ (blue) decays
\label{timeasymmfig}}
\end{figure}

 
 
 The great {\em practical} value of the EPR correlation is instrumental for another consideration as well, namely how to 
 see directly from the data that CP violation is matched by T violation. Fig.\ref{timeasymmfig} shows two distributions, one for the 
 interval  $\Delta t$ between the times of decays $B_d \to l^{+}X$ and $\bar B_d \to \psi K_S$ and the other one for the 
 CP conjugate process $\bar B_d \to l^{-}X$ and $B_d \to \psi K_S$. They are clearly different proving that CP is broken. Yet they show more: the shape of the two distributions is actually the same the 
 only difference being that the average of $\Delta t$ is {\em positive} for $(l^-X)_{\bar B} (\psi K_S)$ and 
 {\em negative} for $(l^+X)_{B} (\psi K_S)$ events. I.e., there is a (slight) preference for $B_d \to \psi K_S$ 
 [$\bar B_d \to \psi K_S$] to occur {\em after} [{\em before}] and thus more [less] slowly (rather than just more rarely) than $\bar B \to l^-X$ [$B \to l^+ X$]. Invoking CPT invariance merely for semileptonic $B$ decays -- yet not for nonleptonic 
 transitions -- synchronizes the $B$ and $\bar B$ decay `clocks'. We thus see that CP and T violation are `just' 
 different sides of the same coin. 
 As explained above, EPR correlations are essential for 
 this argument! 
 
 The reader can be forgiven for feeling that this argument is of academic interest only, since CPT invariance of all 
 processes is based on very general arguments. Yet the main point to be noted is that EPR correlations, which 
 represent some of quantum mechanics' most puzzling features, serve as an essential precision tool, which is routinely used in these measurements. I feel it is thus inappropriate and misleading to refer to EPR correlations as a paradox.

 \subsection{Direct CP Violation and "Yesterday's Sensation, Today's Calibration ..."}
 \label{CALIB}
 
 After the discovery of $K_L \to \pi \pi$ it took 35 years to observe and confirm the existence also of {\em direct} CP violation in the kaon sector.  The analogous development took much less time  in the beauty sector. Direct CP violation has been established in 2004 by both BABAR and BELLE; 
 averaging their results yields:  
 \be 
 A_{CP}^{K\pi} \equiv \frac{\Gamma (\bar B^0 \to K^- \pi^+)- \Gamma (B^0 \to K^+ \pi^-)} 
 {\Gamma (\bar B^0 \to K^- \pi^+)+ \Gamma (B^0 \to K^+ \pi^-)} = - 0.101 \pm 0.025 \pm 0.005 
 \label{DIRECTBDATA}
 \ee
 in amusing agreement with the expectation given in Ref.\cite{JARL2} based on the ansatz of 
 Ref.\cite{SONI}. 
 
It is not widely appreciated that the first strong experimental evidence for direct CP violation had actually emerged  in 
$B_d(t) \to \pi^+\pi^-$. It provides an example of high energy physics' adage "Yesterday's sensation is today's calibration and tomorrow's background". To analyze 
$B_d (t)/\bar B_d(t) \to \pi^+\pi^-$ one again exploits EPR correlations and flavour tagging by a flavour specific decay of the 
other $B$ meson. With the notation 
\be 
R^{+[-]}(\Delta t) \equiv {\rm rate} ((l^{+[-]}X)_B \; {\rm at} \; t; (\pi^+\pi^-)_{B/\bar B} \; {\rm at} \; t+\Delta t)  
\label{RPLUSMINUS}
\ee
one can write the asymmetry between $B_d \to \pi^+\pi^-$ and $\bar B_d \to \pi^+\pi^-$ in terms of two contributions 
distinguishable through their dependance on $\Delta t$: 
\be 
\frac{R^+(\Delta t) - R^-(\Delta t)}{R^+(\Delta t) + R^-(\Delta t)} = S {\rm sin}(\Delta m_B \Delta t) + 
C {\rm cos}(\Delta m_B \Delta t) 
\label{GENFORM}
\ee
with CP invariance requiring $S=0=C$. These coefficients depend on the angles of the CKM triangle, its sides and other hadronic quantities, over 
which theoretical control has not been established yet beyond the general constraint 
$S^2 + C^2 \leq 1$. The status in the summer of 2004 is as follows: 
 \bea 
C &=& \left\{ 
\begin{array}{ll} 
+ 0.58 \pm 0.15  \pm 0.07 & \; \; \; \; {\rm BELLE}  \\ 
- 0.09 \pm 0.15  \pm 0.04 & \; \; \; \; {\rm BABAR} \\ 
 \end{array}  
 \right.  \\ 
 S &=& \left\{ 
\begin{array}{ll} 
- 1.00 \pm  0.21\pm 0.07 & \; \; \; \; {\rm BELLE}  \\ 
- 0.30 \pm 0.17  \pm 0.03 & \; \; \; \; {\rm BABAR} \\ 
 \end{array}  
 \right. \; ; 
 \label{PHI204}
 \eea  
BELLE observes a CP asymmetry with a significance of 5.2 $\sigma$; historically it was the second case of CP violation found in $B$ decays. As explained in the Appendix, in a superweak scenario 
$B_d(t) \to \psi K_S$ provides the calibration; i.e. one would have 
\be 
S = - {\rm sin}2\phi_1 =  - 0.736 \pm 0.049 \; , \;  C = 0 \; . 
\ee
This is ruled out by BELLE's numbers, and {\em direct} CP violation thus established with at least 3.2 $\sigma$ significance, which is significant, but not conclusive yet. No firm conclusions can be derived from the BABAR data at present.

\subsection{The Fly in the CKM Ointment}
\label{FLY}

The observed asymmetry in $B_d \to \psi K_S$ constitutes a striking success for the CKM description and the data 
on $B_d \to \pi \pi$ are quite compatible with it. Yet a potential discrepancy has arisen in $B\to \phi K_S$, 
a channel that had actually been recognized before as having a good potential to reveal physics beyond the SM in general and SUSY in particular. 
In the SM one predicts very confidently:
\be 
C_{\phi K_S} \simeq 0 \; , \;   S_{\phi K_S} \simeq S_{\psi K_S} = 0.736 \pm 0.049 
\ee
In the summer of 2004 the data read as follows:  
 \bea 
C_{\phi K_S} &=& \left\{ 
\begin{array}{ll} 
- 0.08 \pm 0.22  \pm 0.09 & \; \; \; \; {\rm BELLE}  \\ 
+ 0.00 \pm 0.23  \pm 0.05 & \; \; \; \; {\rm BABAR} \\ 
 \end{array}  
 \right.  \\ 
 S_{\phi K_S} &=& \left\{ 
\begin{array}{ll} 
+ 0.06 \pm  0.33\pm 0.09 & \; \; \; \; {\rm BELLE}  \\ 
+ 0.50 \pm 0.25 ^{+0.07}_{-0.04}  \pm 0.03 & \; \; \; \; {\rm BABAR} \\ 
 \end{array}  
 \right. \; .
 \label{PHIKSDATA}
\eea
While BABAR's findings are consistent with the predictions, BELLE's results point to a 2.2 
$\sigma$ discrepancy. This has attracted considerable attention since the CP asymmetries in a whole class of related modes all seem to show a shift relative to the CKM predictions, and this class is expected to exhibit a high sensitivity to the intervention of New Physics.  
It will take some time, though, to clarify the experimental situation.

 \subsection{"... and Tomorrow's Background" -- the Cosmic Connection}
 \label{COSMIC}
 
 As mentioned before the most ambitious goal in CP studies is to understand the observed baryon number of the 
 Universe as a dynamically generated quantity, for which CP violation is one of the three central ingredients. We know now that the standard CKM dynamics while successful in describing CP breaking observed in particle decays 
 is quite incapable to provide this cosmic connection: for its effective strength is too feeble, and it cannot induce the 
 required {\em first} order phase transition at the electroweak symmetry breaking. Thus this program requires the intervention of New Physics, for which many interesting scenarios have been put forward. 
 
Of course one wants to identify manifestations of such hypothetical New Physics in processes that can be probed 
in  reproducible laboratory experiments. In general such new dynamics will affect CP asymmetries in $B$ decays; yet 
those `suffer' from the background of large asymmetries due to CKM dynamics that are present. This poses considerable -- though hopefully not insurmountable -- challenges on the experimental as well as theoretical side.  

The encouraging news is the aforementioned demystification of CP violation: as the example of CKM dynamics shows in general there is no impediment to the CP violating complex phase being large; this will presumably be needed to generate the baryon number of the Universe.

 \subsection{Gateway to a New World: Neutrino Oscillations}
 \label{GATEWAY}
 
  A particular intriguing class of models interprets the baryon number as a secondary effect derived from the primary 
 phenomenon of leptogenesis; i.e., first a non-vanishing lepton number is generated  for the Universe, which is then transmogrified into a baryon number. This provides new impetus -- actually makes it mandatory -- to search for manifestations of CP violation in the dynamics of leptons. Relevant processes are 
 \begin{itemize}
 \item 
 neutrino oscillations, 
 \item 
 the decays of $\tau$ (and $\mu$) leptons 
 \footnote{The transition $\tau ^- \to \mu^- \mu^+ \mu^-$ forbidden in the SM is the leptonic analogue of the quark transition $b\to s \bar s s$ driving $B \to \phi K_S$.} 
 and 
 \item 
 atomic electric dipole moments. 
 \end{itemize}
 
 The observation of neutrino oscillations through solar and `atmospheric' neutrinos  -- which constitutes the third 
 column of the `phase transition' at the turn of the millenium referred to before -- opens up a new world to probe 
 fundamental physics in general and CP violation in particular, which is of direct relevance for leptogenesis. 
 
 Neutrino oscillations can occur only when the different neutrino types are {\em not} mass degenerate. The 
 leptonic charged current couplings are then described by the so-called PMNS matrix \cite{PMNS} in analogy to the 
 CKM matrix for quarks. The $3\times 3$ PMNS matrix is far from exhibiting the strictly hierarchical form of 
 Eq.(\ref{WOLFKM}) -- yet that is not surprising since there is a fundamental distinction between charged and neutral 
 fermion fields: for the latter can acquire a Majorana mass term in addition to a Dirac mass; through the so-called 
 see-saw mechanism \cite{SEESAW,CPBOOK} this can provide a natural explanation why the neutrino masses are so tiny compared to all other fermion masses. The fact that the form of the CKM and PMNS matrices does not exhibit a unified pattern should therefore not be seen as a drawback -- on the contrary! This can be illustrated by the following true anecdote: a long time ago a French politician was asked whether his opposition to German unification does not reveal his basic dislike of Germany. He rejected this assertion by saying that he truly loves Germany and he is therefore overjoyed that there are two Germanies he can love \footnote{There is another lesson to be learnt from this analogy: even when unification seems impossible, it can happen in due course.}.

 Irrespective of that connection  finding CP violation in leptodynamics would complete the `de-mystification' of CP violation that, as repeatedly mentioned before, has occurred in quark dynamics.

 \section{Reflections about the Past and the Future}
 \label{SUMMARY}
 
 After a long gestation several important developments have come to fruition starting in 1999: 
 \begin{itemize}
 \item 
 A second qualitatively different source of CP violation, namely direct CP violation, has unequivocally been established 
 experimentally in $K_L \to \pi^+\pi^-$ vs. $K_L \to \pi^0\pi^0$ decays. 
 \item 
 Data together with truly {\em minimal} theoretical assumptions confirm that the CP asymmetry observed in neutral kaon decays is fully matched by violation of time reversal invariance T as required by CPT invariance. 
 \item 
 The large CP asymmetry predicted for $B_d \to \psi K_S$ by the CKM description in the old-fashioned sense -- i.e., 
 the prediction was made well before the experimental findings were known -- has been confirmed by the data to an amazing degree. 
 \item 
 The measurements show -- again with minimal assumptions -- that the observed asymmetry is matched by a commensurate violation of T invariance. 
 \item 
 A CP asymmetry has been observed in a second class of channels, namely $B_d \to \pi^+\pi^-$, yet its detailed 
 interpretation is still open to debate. 
 \item 
 EPR correlations provide an indispensable {\em precision} tool for the experimental analyses;  these are actually a novel type of EPR correlations, where the two correlated states change their identity on a time scale of picoseconds! The data are 
 fully consistent with quantum mechanics' specific predictions. 
 \item 
 With all these observations (and others before) consistent with predictions and expectations based on the CKM description, the latter has been promoted from an ansatz to a tested theory. 
 \item 
 This progress was achieved through an intimate interplay between theoretical suggestions, experimental results and novel concepts in detector and accelerator design.

 \end{itemize}
 
 The experimenters will not rest on their laurels, however well-deserved that would be. The $B$ factories at KEK and SLAC will run with ever increasing statistics and refined experimental techniques to probe $B$, charm and 
 $\tau$ decays with higher and higher sensitivity; hopefully one of them will be upgraded to a `Super-B' factory 
 \cite{SUPERB}. In a few years they will be joined and pushed by experiments performed at the hadronic colliders of Fermilab in the US and CERN in Europe. Such further efforts are actually mandatory. For the presently achieved successes do not resolve central mysteries of the SM: Why are there families of quarks and leptons? Why three -- or are there more? Why has the CKM matrix its highly unusual hierarchical structure? Why does QCD conserve CP invariance in flavour diagonal transitions to such a high degree of precision? 
 
 In addition to these indirect arguments for the incompleteness of the SM, there are more direct ones as well: 
 \begin{itemize}
 \item 
 At present there are intriguing indications from BELLE's measurements that the CP asymmetry in $B_d \to \phi K_S$ exhibits a stunning deviation from the CKM prediction. 
 \item 
 What is the nature of the new dynamics needed to generate the observed matter-antimatter asymmetry in the Universe?
 \item 
 What are the forces driving the observed neutrino oscillations? Will they exhibit CP violation as well? 
 
 \end{itemize}
 Continuing dedicated and comprehensive studies should provide us with information that will facilitate our searches 
 for answers to these fundamental questions.  I consider further experimental information as crucial in this endeavour, 
 since it will point us in the right direction -- yet it will not be sufficient: it will be nontrivial to digest the experimental information theoretically.

\noindent {\bf Epilogue} 

\noindent The physicists' tale of CP violation is a profound one that is teaching us important lessons that go beyond identifying Nature's fundamental forces. It has lead us to formulate questions about `Nature's Grand Design' that we did 
not think about at the beginning of our journey: Is it really possible to create a Universe  with only matter, but no domains of antimatter? What about other Universes that might exist besides ours? Maybe CP violation and the whole family structure embedded into the SM carries a coded message about extra dimensions beyond the well-known 1+3 time-space dimensions. What about the very structure of time? What made it one-dimensional -- or are there additional though hidden dimensions of time. What about the arrow of time in such exotic scenarios? 

These are questions, where we still have not even clues about the answers. Yet this is probably quite appropriate for the subjects discussed at this `International Colloquium on the Science of Time'. 

A final observation: for me it reflects one of the noblest and thus most encouraging features of the human race that there is a continuous stream of young people eager to commit themselves to exploring Nature for the gain of knowledge for knowledge's sake and that they bring to the table unusual amounts of talent, dedication, persistence and creativity. 

\renewcommand{\theequation}{A.\arabic{equation}}
\renewcommand{\thesection}{}
\setcounter{equation}{0}
\setcounter{section}{0}
\section{Appendices}
\renewcommand{\thesubsection}{A.\arabic{subsection}}

\subsection{On {\em direct} CP Violation} 

It is often said -- or at least implied -- that $S \neq 0$ and $C \neq 0$ reflect two distinct sources 
of CP violation. Indeed $C \neq 0$ reveals unequivocally {\em direct} CP violation. Yet the situation with $S\neq 0$ is more complex,  as can be read off from the explicit expression for $S$: 
\be 
S = {\rm Im} \frac{q}{p} \frac{T(\bar B_d \to \pi^+ \pi^-)}{T(B_d \to \pi^+ \pi^-)}
\ee
$\frac{q}{p}$ reflects $\Delta B=2$ dynamics driving $B^0 - \bar B^0$ oscillations, and its phase provides a measure for {\em indirect} CP violation; yet the ratio of the instantaneous transition amplitudes $T(\bar B_d [B_d]\to \pi^+ \pi^-)$ represents $\Delta B =1$ dynamics, including 
their CP features. As a further complication the phases of $\frac{q}{p}$ and 
$T(\bar B_d \to \pi^+ \pi^-)/T(B_d \to \pi^+ \pi^-)$ depend on the phase {\em convention} adopted for the definition of $\bar B_d$ -- only their product does not. Therefore as long as CP violation is studied in a {\em single} channel, it is a matter of convention whether $S\neq 0$ is called an indirect or a direct CP violation. However once one can compare it in two final states common to $B_d$ and $\bar B_d$ decays -- like in $B_d(t) \to \psi K_S$ vs.  
$B_d (t) \to \pi^+ \pi^-$ -- the two cases can be distinguished. For if there is {\em no direct} 
CP violation -- i.e. for a superweak scenario -- one has 
\be 
S^{\pi \pi} = - S^{\psi K_S} \; , 
\ee 
where the minus sign is due to the final states $\pi \pi$ and $\psi K_S$ having opposite CP parity. 
Finding instead 
\be 
S^{\pi \pi} \neq - S^{\psi K_S}
\ee
establishes unequivocally the intervention of direct CP violation, since it shows there is a relative phase between $qT(\bar B_d \to \pi^+ \pi^-)/pT(B_d \to \pi^+ \pi^-)$ and 
$qT(\bar B_d \to\psi K_S)/pT(B_d \to \psi K_S)$ and thus also between 
$T(\bar B_d \to \pi^+ \pi^-)/T(B_d \to \pi^+ \pi^-)$ and 
$T(\bar B_d \to\psi K_S)/T(B_d \to \psi K_S)$, i.e. in pure $\Delta B=1$ amplitudes. Likewise for 
\be 
S^{\phi K_S} \neq S^{\psi K_S} \; {\rm or} \; S^{\eta K_S} \neq  S^{\psi K_S} \; 
{\rm or} \; S^{K_S \pi^0} \neq S^{\psi K_S} \; {\rm etc.}
\ee
One should also note that such direct CP violation might {\em not} generate $C\neq 0$, since it does not require the presence of two different amplitudes with a nontrivial phase shift between them. 

\subsection{A New Opening for `Patience'}
\label{PATIENCE}

The CDF experiment at FNAL has obtained the intriguing, though preliminary result that the two 
$B_s$ mass eigenstates might possess significantly different widths \cite{CDFBS}: 
\be 
\frac{\Delta \Gamma (B_s)}{\Gamma (B_s)} = 0.65 ^{+0.25}_{-0.33} \pm 0.01 \; . 
\label{DELTAGAMMAS}
\ee
If the true number is close to this central value, which is about four times larger than predicted, then history could repeat itself. For in qualitative analogy to the $K_L$ - $K_S$ case 
`patience' would be awarded; i.e., an initial beam of $B_s$ and $\bar B_s$ mesons 
would turn itself into an increasingly pure beam of the long lived meson, since the short lived one would decay away faster. This would open the door for novel searches for CP violation in 
$B_s$ decays, where an inability to resolve the fast $B_s - \bar B_s$ oscillations driven by 
$\Delta M(B_s)$ might turn into a virtue rather than a vice. The fact that this could be achieved 
when the short and long lived components have lifetimes of close to one and two picoseconds rather than $\sim 0.1$ and $\sim 50$ nanoseconds, as it was the case for $K_S$ and $K_L$,  
exemplifies the impressive progress in detector technology.



\begin{thebibliography}{99}

\bibitem{EPR}
A. Einstein, B. Podolsky, N. Rosen, {\em Phys.Rev.} {\bf 47} (1935) 777. 

\bibitem{FITCH} 
J.H. Christensen {\em et al.}, {\em Phys.Rev.Lett.} {\bf 13} (1964) 138. 

\bibitem{CPBOOK} 
A very detailed discussion of all aspects of CP violation can be found in: 
I.I. Bigi, A.I. Sanda, `CP Violation', Cambridge Monographs on Particle Physics, Nuclear Physics and Cosmology, 
Cambridge University Press, 2000. 

\bibitem{SAKH}
A.D. Sakharov, {\em JETP Lett.} {\bf 5} (1967) 24; for an updated review see: A.D. Dolgov, hep-ph/9707419.  

\bibitem{KRAMERS}
H. A. Kramers, {\em Proc. Acad. Sci. Amsterdam} {\bf 33} (1930) 959; see also: F.J. Dyson, 
{\em J. Math. Phys.} {\bf 3} (1962) 140. 

\bibitem{ROOS} 
B. Laurent, M. Roos, {\em Phys.Lett.} {\bf 13} (1964) 269; {\em ibid.} {\bf 15} 104.

\bibitem{BLUCHER} 
E. Blucher, these Proceed.


\bibitem{ZVARTANIK}
D. Zavrtanik, these Proceed. 

\bibitem{SUPERWEAK}
L. Wolfenstein, {\em Phys. Rev. Lett.} {\bf 13} (1964) 562. 

\bibitem{KM} 
M. Kobayashi, T. Maskawa, {\em Prog. Theor. Phys.} {\bf 49} (1973) 652; 
N. Cabibbo, {\em Phys. Rev. Lett.} {\bf 10} (1963) 531. 


\bibitem{MOHA}
R. Mohapatra, {\em Phys. Rev.} {\bf D6} (1972) 2023. 




\bibitem{NIU} 
K. Niu, E. Mikumo, Y. Maeda,  {\em Prog. Theor. Phys.} {\bf 46} (1971) 1644. 

\bibitem{JARL}
C. Jarlskog in: {\em CP Violation}, ed. C Jarlskog
(World Scientific, Singapore, 1988).

\bibitem{BS1} 
A.B. Carter, A.I. Sanda, {\em Phys. Rev.} {\bf D23} (1981) 1567; 
I.I. Bigi, A.I. Sanda, {\em Nucl. Phys.} {\bf B193} (1981) 85. 

\bibitem{BS2} 
I.I. Bigi, A.I. Sanda, {\em Nucl. Phys.} {\bf B281} (1987) 41. 

\bibitem{OPAL} 
OPAL Collab., K. Ackerstaff {\em et al.}, {\em Eur. Phys. J.} {\bf C5} (1998) 379. 

\bibitem{CDF}
CDF Collab., {\em Phys. Rev.} {\bf D61} (2000) 072005. 
       
\bibitem{ODDONE}
P. Oddone, in: Proceed. of the 1987 UCLA Workshop on the Linear Collider $B\bar B$ Factory Conceptual 
Design, Los Angeles, 1987.  

\bibitem{EARLY}
T. Morozumi, C.S. Lim, A.I. Sanda, {\em Phys. Rev. Lett.} {\bf 65} (1990) 404. 

\bibitem{TRIESTE} 
S. Bertolini, J. Eeg, M. Fabbrichesi, {\em Rev. Mod. Phys.} {\bf 72} (2000) 65. 

\bibitem{BEFORETOP} 
I.I. Bigi, in: Proceed. of `Les Rencontres de Physique de la Vallee d'Aoste, La Thuile, Italy, 1991; 
in: Proceed. of `Les Rencontres de Moriond, Les Arcs, France, 1992. 

\bibitem{PRECISE}  
F. Parodi, P. Roudeau, A. Stocchi, {\em Nuovo Cim.} {\bf  A112} (1999) 833. 

\bibitem{SONI} 
M. Bander, D. Silverman, A. Soni, {\em Phys.Rev.Lett.} {\bf 43} (1979) 242. 

\bibitem{JARL2} 
I.I. Bigi, V.A. Khoze, N.G. Uraltsev, A.I. Sanda, in: {\em CP Violation}, ed. C Jarlskog
(World Scientific, Singapore, 1988), p. 218.

\bibitem{KTEV}
KTeV Collab., A. Alavi-Harati {\em et al.}, {\em Phys. Rev.} {\bf D67} (2003) 012005. 


\bibitem{NA48} 
NA48 Collab., A. Lai {\em et al.}, {\em Eur. Phys. J.} {\bf C22} (2001) 231.

\bibitem{CDFBS} 
http://www-cdf.fnal.gov/physics/new/bottom/040708.blessed-dgog-bsjpsiphi/
      

\bibitem{PMNS}
Z. Maki, M. Nakagawa, S. Sakata, {\em Prog. Theor. Phys.} {\bf 30} (1963) 727; 
B. Pontecorvo, {\em J. Exp. Theor. Phys.} {\bf 33} (1957) 549.

\bibitem{SEESAW}
M. Gell-Mann, R. Slansky, P. Ramond, in: {\em Supergravity}, North Holland, 1979, p. 315; 
T. Yanagida, in: {\em Proc. Workshop on Unified Theory and Baryon Number in the Universe}, KEK, 
Japan, 1979. 

\bibitem{SUPERB}
I.I. Bigi, A.I. Sanda, hep-ph/0401003. 





\end{thebibliography}
\end{document}